%% file: main.tex
\documentclass[conference]{IEEEtran}
\IEEEoverridecommandlockouts
\usepackage{cite}
\usepackage{amsmath,amssymb,amsfonts}
\usepackage{algorithmic}
\usepackage{graphicx}
\usepackage{textcomp}
\usepackage{xcolor}
\usepackage[capitalise]{cleveref}

\def\BibTeX{{\rm B\kern-.05em{\sc i\kern-.025em b}\kern-.08em
    T\kern-.1667em\lower.7ex\hbox{E}\kern-.125emX}}

\begin{document}

\title{Multi-Channel FFT Architectures Designed via Folding and Interleaving
\thanks{This research was supported in part by the National Science Foundation under grant number CCF-1954749.}
}
\author{\IEEEauthorblockN{Nanda K. Unnikrishnan}
\IEEEauthorblockA{\textit{Dept. Electrical and Computer Engineering} \\
\textit{University of Minnesota}\\
Minneapolis MN, USA  \\
unnik005@umn.edu}
\and
\IEEEauthorblockN{Keshab K. Parhi}
\IEEEauthorblockA{\textit{Dept. Electrical and Computer Engineering} \\
\textit{University of Minnesota}\\
Minneapolis MN, USA  \\
parhi@umn.edu}
}
\IEEEaftertitletext{\vspace{-1.5\baselineskip}}
\maketitle

\input{sections/S0_abstract}
\input{sections/S1_Introduction}
\input{sections/S3_proposed}
\input{sections/S4_comparisons}
\input{sections/S5_conclusion}

\bibliographystyle{IEEEtran}
\bibliography{references}

\end{document}

%% file: sections/S0_abstract.tex
\begin{abstract}
Computing the FFT of a single channel is well understood in the literature. However, computing the FFT of multiple channels in a systematic manner has not been fully addressed. This paper presents a framework to design a family of multi-channel FFT architectures using {\em folding} and {\em interleaving}. Three distinct multi-channel FFT architectures are presented in this paper. These architectures differ in the input and output preprocessing steps and are based on different folding sets, i.e., different orders of execution.
\end{abstract}


%% file: sections/S1_Introduction.tex

\section{Introduction}
Fast Fourier Transform (FFT) algorithm is a critical part of modern signal processing and machine learning systems, and is used in applications ranging from digital communication~\cite{MIMO_FFT_comm_1,MIMO_FFT_2,MIMO_FFT_3,MIMO_FFT_4} to generating features for neural networks.  There has been significant research on design of pipelined \cite{PipelinedFFT12,fft_survey2021, pipelined_1} and parallel \cite{mixed_radix_parallel, parallel_2} FFT architectures for both complex and real-valued input signals \cite{real_FFT1, real_FFT2, real_FFT3}. These architectures were designed mostly with systematic principles. However, systematic design of architectures for multi-channel has not been fully explored. 

We can broadly classify existing multi-channel approaches into two categories. The first approach is memory-based which store all input channels into a large memory bank \cite{MIMO_FFT_4, MIMO_FFT_5, MIMO_FFT_6}. While this approach has less data movement and reordering, it requires a significantly larger memory footprint, especially simultaneous writing and reading support. The second approach uses a combination of delay elements and switches to create data commutators \cite{MIMO_FFT_comm_1, MIMO_FFT_comm_2,MIMO_FFT_comm_3}. These data commutators can manipulate input data into the proper format to interleave the channels and handle the data reordering for the first stage. While this approach is memory efficient, it lacks systematic exploration and is tailored for specific hardware applications or architectures. Therefore, this paper explores the systematic design of multi-channel FFT architectures by applying folding and interleaving to existing architectures. 


The rest of the paper is organized as follows. Section II describes the proposed interleaving models and how to derive them. Section III compares the different proposed approaches and their advantages over existing designs. Finally, section IV summarizes the main conclusions of the paper. 

%% file: sections/S3_proposed.tex
\begin{figure}
    \centering
    \includegraphics[width=0.95\linewidth]{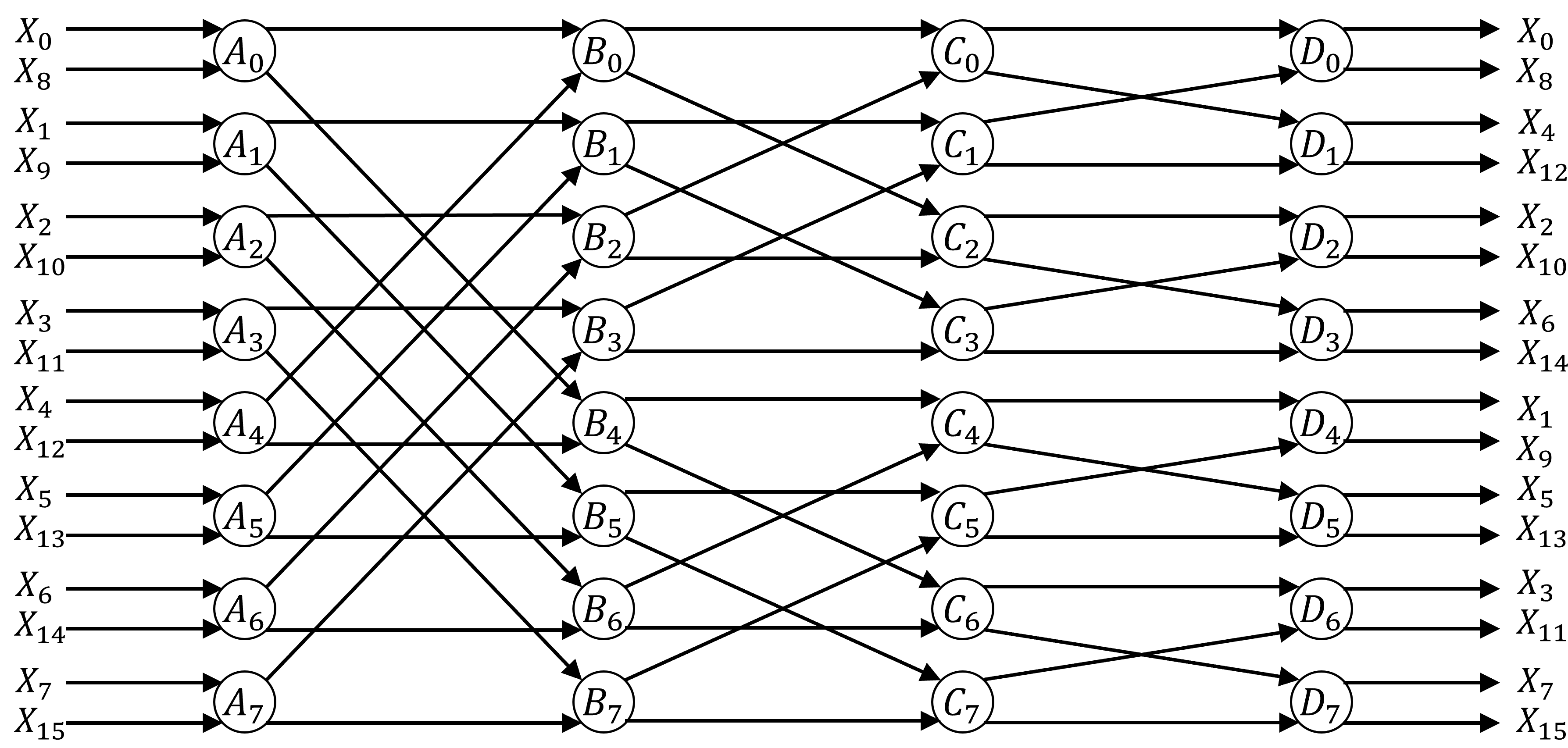}
    \caption{DFG of a Radix-2 16-point DIF FFT with processors allocation for folding.}
    \label{fig:DIF-DFG}
\end{figure}

\section{Proposed multi-channel FFT architectures}
This section presents the multi-channel interleaved architectures for complex-valued signals with the radix-2 algorithm. These architectures were designed with the use of folding sets \cite{folding}. Folding sets are ordered sets that describe how operations map to a hardware resource in a time-multiplexed manner. Through folding, we can derive different FFT architectures by varying the operations' order. Additionally, we can use the process of interleaving \cite{interleaving} to alternate between the computation of each channel. Using folding sets, we present three architectures for a multi-channel FFT. For brevity, we only present the folding sets and the final architectures.

\begin{figure}
    \centering
    \includegraphics[width=0.80\linewidth]{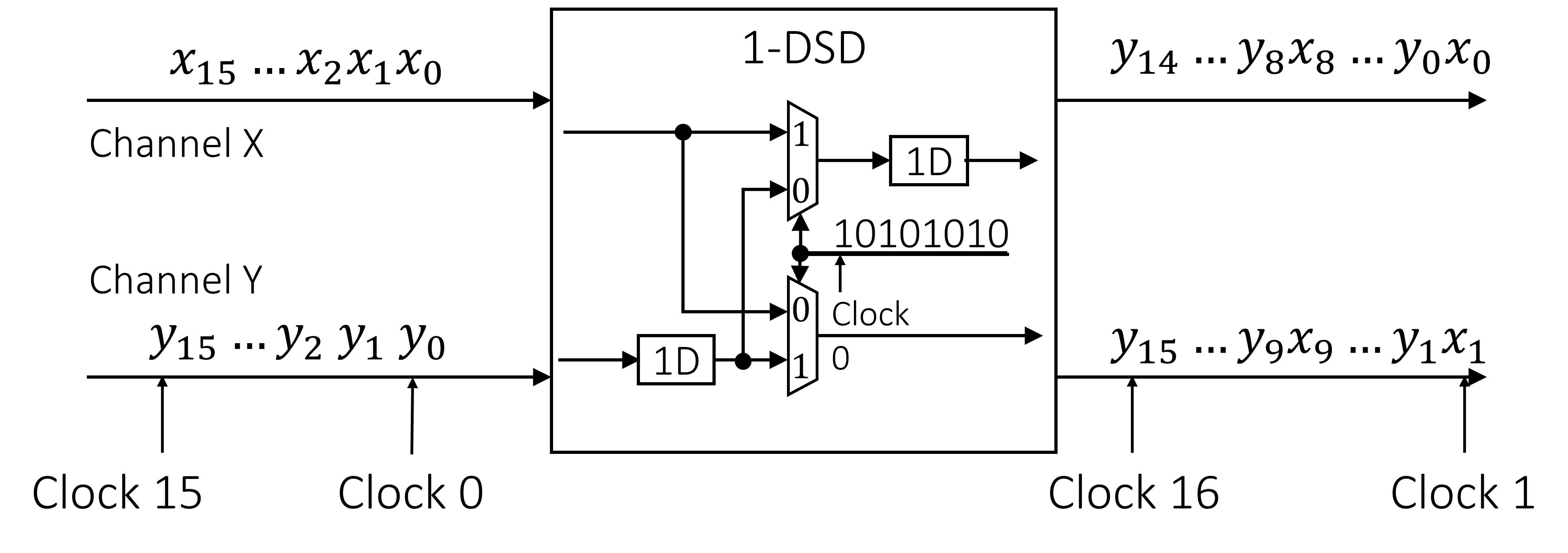}
    \caption{Pre-processing multi-channel input into an interleaved 2-parallel form using a 1-DSD circuit.}
    \label{fig:1DSD}
\end{figure}

\begin{figure*}
    \centering
    \includegraphics[width=0.79\linewidth]{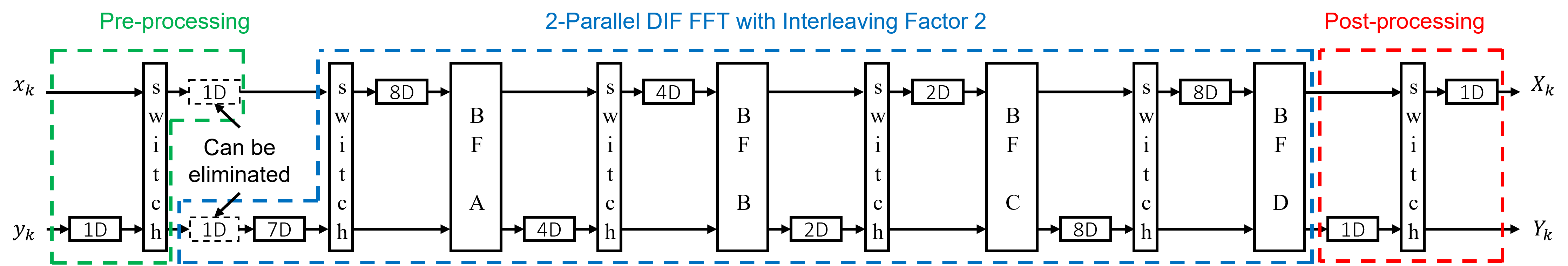}
    \caption{Proposed 2-parallel interleaved architecture (Architecture 1) for the computation of a 2 channel 16-point DIF FFT.}
    \label{fig:2-parallel-fft-inter}
\end{figure*}

\subsection{Architecture 1: 2-parallel FFT architecture with interleaving factor 2}

A direct approach to multi-channel interleaving is to take existing FFT architectures and perform the interleaving operation. There are multiple candidates for the choice of a base architecture \cite{2-parallel}, and for this example, we use the 16-point 2-parallel DIF FFT proposed in Fig. 12 of \cite{PipelinedFFT12} as the starting point for interleaving. The base architecture defines the folding sets in \cref{eq:FS} for the dataflow graph shown in \cref{fig:DIF-DFG}.

Interleaving by a factor of 2 doubles the number of delays in the architecture. In terms of the folding set, we insert null operations in the new locations as shown in \cref{eq:FS_double}. This change leads to a 2-parallel FFT architecture that accepts inputs on alternate cycle. The system still maintains the throughput of a single channel system, albeit with $50\%$ utilization.

{\scriptsize
\begin{align}
    A & = \left\{A_0, A_2, A_4, A_6, A_1, A_3, A_5, A_7\right\} \nonumber\\
    B & = \left\{B_5, B_7, B_0, B_2, B_4, B_6, B_1, B_3\right\}\nonumber\\
    C & = \left\{C_3, C_5, C_7, C_0, C_2, C_4, C_6, C_1\right\}\nonumber\\
    D & = \left\{D_2, D_4, D_6, D_1, D_3, D_5, D_7, D_0\right\}\label{eq:FS}
\end{align}
\begin{align}
    A & = \left\{A_0, \emptyset, A_2, \emptyset, A_4, \emptyset, A_6, \emptyset, A_1, \emptyset, A_3, \emptyset, A_5, \emptyset, A_7, \emptyset\right\} \nonumber\\
    B & = \left\{B_5, \emptyset, B_7, \emptyset, B_0, \emptyset, B_2, \emptyset, B_4, \emptyset, B_6, \emptyset, B_1, \emptyset, B_3, \emptyset\right\}\nonumber\\
    C & = \left\{C_3, \emptyset, C_5, \emptyset, C_7, \emptyset, C_0, \emptyset, C_2, \emptyset, C_4, \emptyset, C_6, \emptyset, C_1, \emptyset\right\}\nonumber\\
    D & = \left\{D_2, \emptyset, D_4, \emptyset, D_6, \emptyset, D_1, \emptyset, D_3, \emptyset, D_5, \emptyset, D_7, \emptyset, D_0, \emptyset\right\}\label{eq:FS_double}
\end{align} }
We can use these null operations to interleave the second FFT channel to the same hardware as shown in \cref{eq:FS_inter}, where prime operations refer to the second channel. 

{\scriptsize
\begin{align}
    A = & \left\{A_0, A'_0, A_2, A'_2, A_4, A'_4, A_6, A'_6,  A_1, A'_1, A_3, A'_3, A_5, A'_5, A_7, A'_7 \right\}\nonumber\\
    B = & \left\{B_5, B'_5, B_7, B'_7, B_0, B'_0, B_2, B'_2, B_4, B'_4, B_6, B'_6, B_1, B'_1, B_3, B'_3\right\}\nonumber\\
    C = & \left\{C_3, C'_3, C_5, C'_5, C_7, C'_7, C_0, C'_0, C_2, C'_2, C_4, C'_4, C_6, C'_6, C_1, C'_1\right\}\nonumber\\
    D = & \left\{D_2, D'_2, D_4, D'_4, D_6, D'_6, D_1, D'_1, D_3, D'_3, D_5, D'_5, D_7, D'_7, D_0, D'_0\right\}\label{eq:FS_inter}
\end{align}
}
\cref{fig:2-parallel-fft-inter} describes the proposed architecture for the 2-parallel DIF FFT interleaved by a factor of 2. Using interleaving, we can derive this architecture by doubling the number of delays in the original circuit and interleaving the input signals. For example, the above folding set alternately accepts two inputs from each channel. To pre-process the data into this format, we make use of a one delay-switch-delay (1-DSD) shown in \cref{fig:1DSD}. The same 1-DSD circuit can be used at the final output to post-process the information such that the corresponding results of each channel line up. This ordering is advantageous if we need to multiply the related FFT results of the two channels. In addition, there is a delay in both the upper and lower path after the pre-processing step. We can eliminate these delays from the circuit without any consequence by the principle of reverse pipelining~\cite{parhi_book}. If we require the outputs in the natural order, then we must implement a reorder circuit (REOC) in \cite{PipelinedFFT12} with seven registers per channel or seven registers in total if we multiply and use each channel.

\begin{figure}
    \centering
    \includegraphics[width=0.75\linewidth]{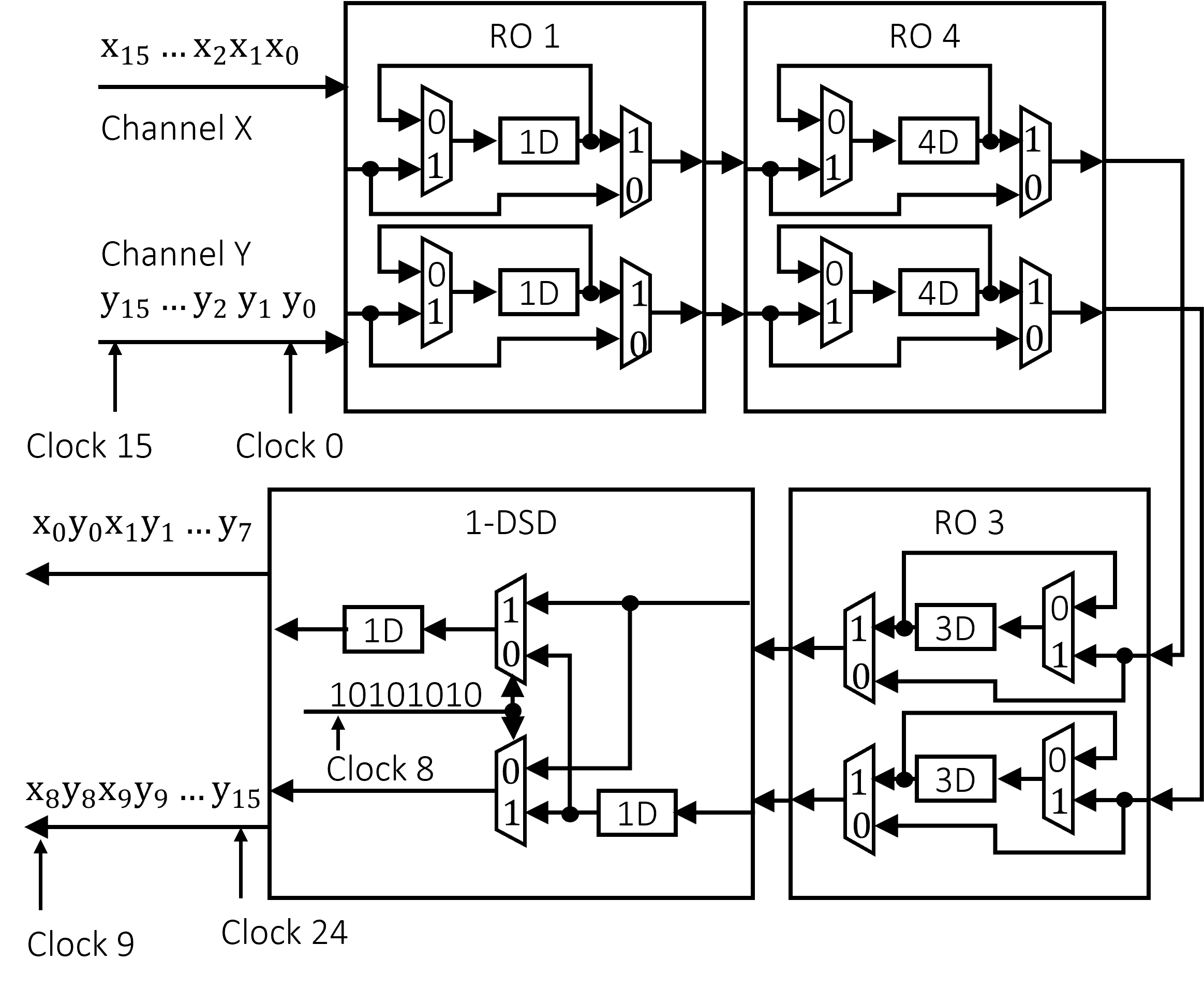}
    \caption{Pre-processing multi-channel input into an alternative interleaved 2-parallel form.}
    \label{fig:inter-pre-2}
\end{figure}

\begin{figure}
    \centering
    \includegraphics[width=0.50\linewidth]{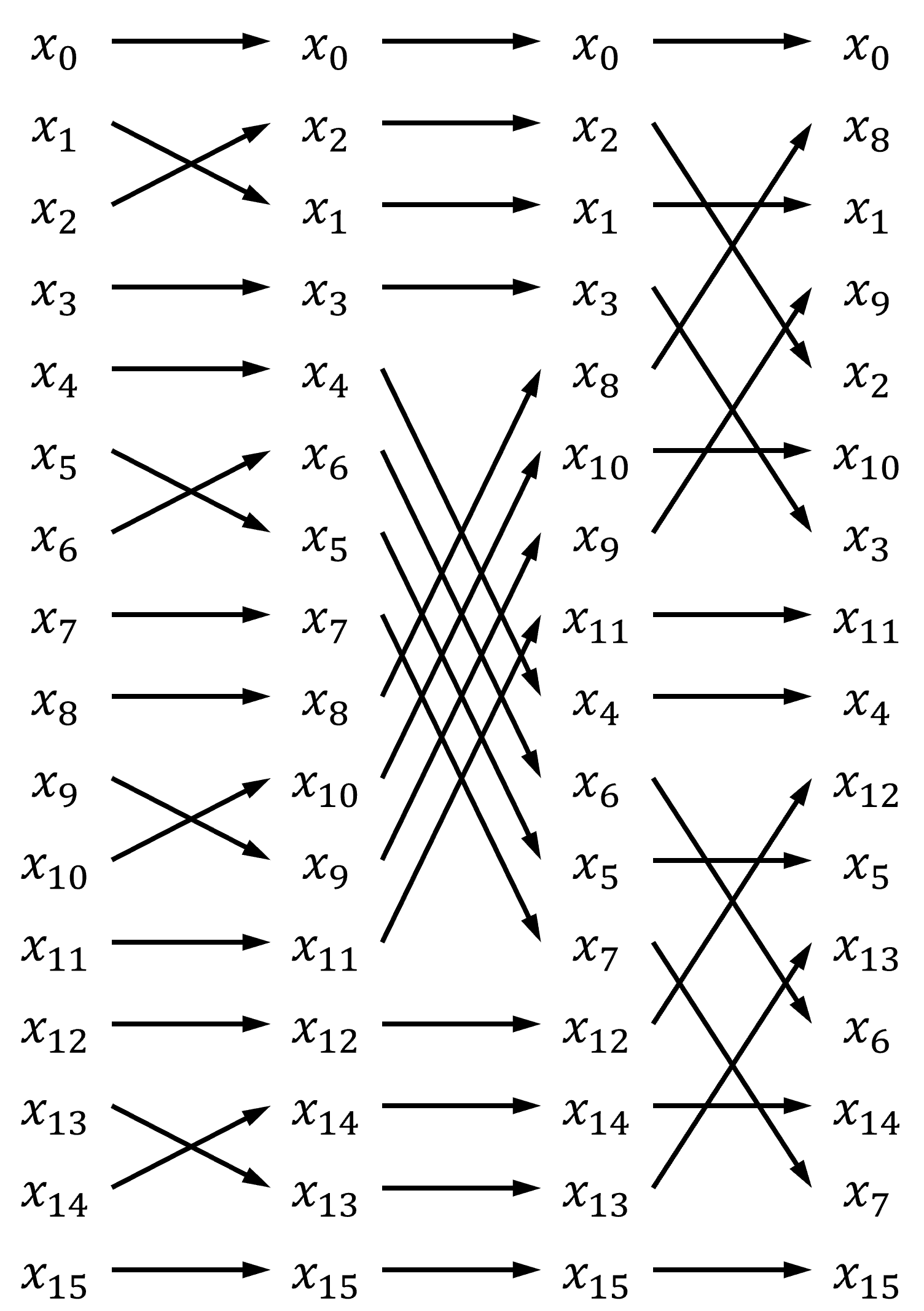}
    \caption{Data movement in the reorder circuits.}
    \label{fig:inter-pre-2-RO}
\end{figure}

\begin{figure*}
    \centering
    \includegraphics[width=0.80\linewidth]{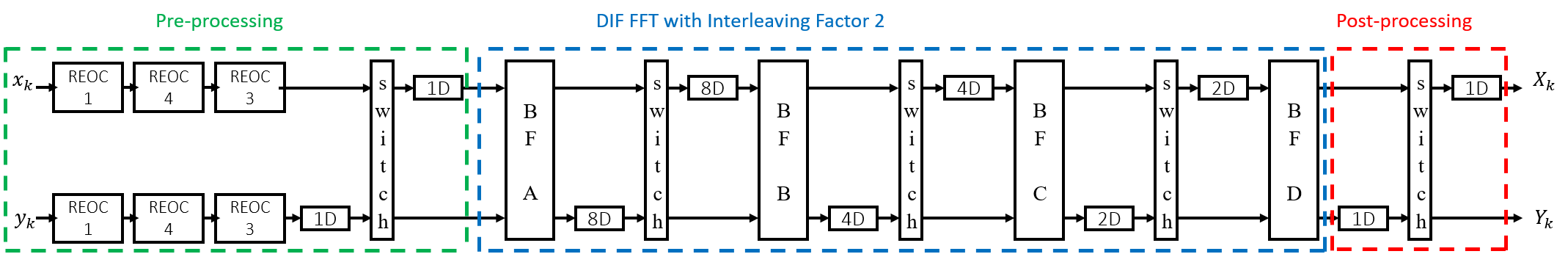}
    \caption{Proposed interleaved architecture (Architecture 2) for the computation of a 2-channel 16-point DIF FFT.}
    \label{fig:fft-inter-basic}
\end{figure*} 
\begin{figure*}
    \centering
    \includegraphics[width=0.80\linewidth]{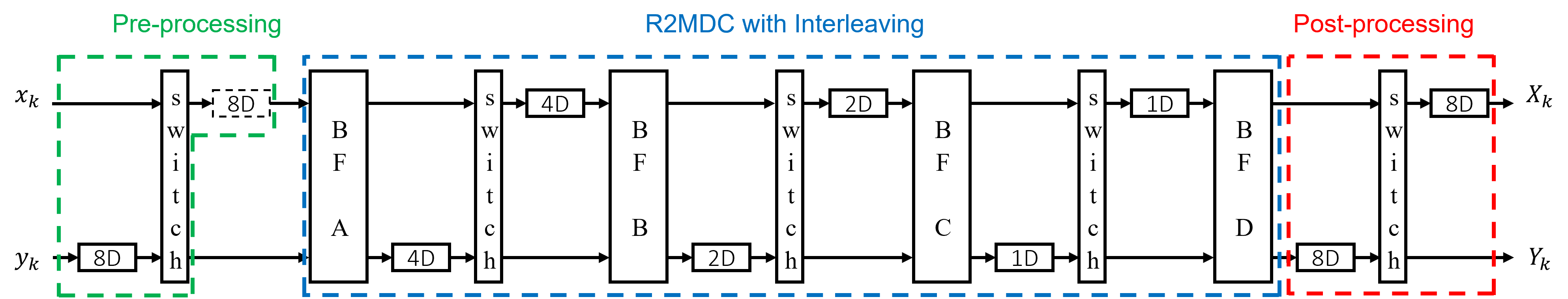}
    \caption{Proposed interleaved architecture (Architecture 3) based on R2MDC for the computation of a 2-channel 16-point DIF FFT.}
    \label{fig:fft_inter_comm}
\end{figure*}

\subsection{Architecture 2: Alternate 2-parallel FFT architecture with interleaving factor 2}

The main difference between~\cref{eq:FS_inter_2} and~\cref{eq:FS_inter} is the ordering of the computations within the folding set. \cref{eq:FS_inter_2} applies a simple ordering scheme for mapping the computations to the processor. As a consequence, the output order of the circuit follows the output order of the DFG in~\cref{fig:DIF-DFG}. To use the new folding sets, we design a pre-processing step, shown in \cref{fig:inter-pre-2}, that takes the two input channels and generates a sequence required for the circuit. 

The circuit in \cref{fig:inter-pre-2} uses REOC circuits to swap data entries that are a fixed delay apart. For this architecture, the data movement function of the REOC circuit is shown in \cref{fig:inter-pre-2-RO}. The first step in the pre-processing uses a combination of  $RO1+RO3+RO4$ to bring all the inputs used for a butterfly to adjacent cycles. Additionally, the circuit ensures that the order of operations matches the order in the DFG in \cref{fig:DIF-DFG}. Finally, the post-processing step uses a 1-DSD to un-interleave the two FFT channel outputs. The overall architecture showing the pre-processing, post-processing, and FFT modules is shown in~\cref{fig:fft-inter-basic}. The outputs of the circuit are in a standard bit-reversed order, and we use bit reversal circuits like~\cite{garrido_bitreverse} to bring them to a natural order. This reversal circuit corresponds to 9 registers per output channel for a 16 point FFT. 

{\scriptsize
\begin{align}
    A = & \left\{A_0, A'_0, A_1, A'_1, A_2, A'_2, A_3, A'_3, A_4, A'_4, A_5, A'_5, A_6, A'_6, A_7, A'_7 \right\}\nonumber\\
    B = & \left\{B_4, B'_4, B_5, B'_5, B_6, B'_6, B_7, B'_7, B_0, B'_0, B_1, B'_1, B_2, B'_2, B_3, B'_3\right\}\nonumber\\
    C = & \left\{C_2, C'_2, C_3, C'_3, C_4, C'_4, C_5, C'_5, C_6, C'_6, C_7, C'_7, C_0, C'_0, C_1, C'_1\right\}\nonumber\\
    D = & \left\{D_1, D'_1, D_2, D'_2, D_3, D'_3, D_4, D'_4, D_5, D'_5, D_6, D'_6, D_7, D'_7, D_0, D'_0\right\}\label{eq:FS_inter_2}
\end{align}
}
\begin{figure}
    \centering
    \includegraphics[width=0.8\linewidth]{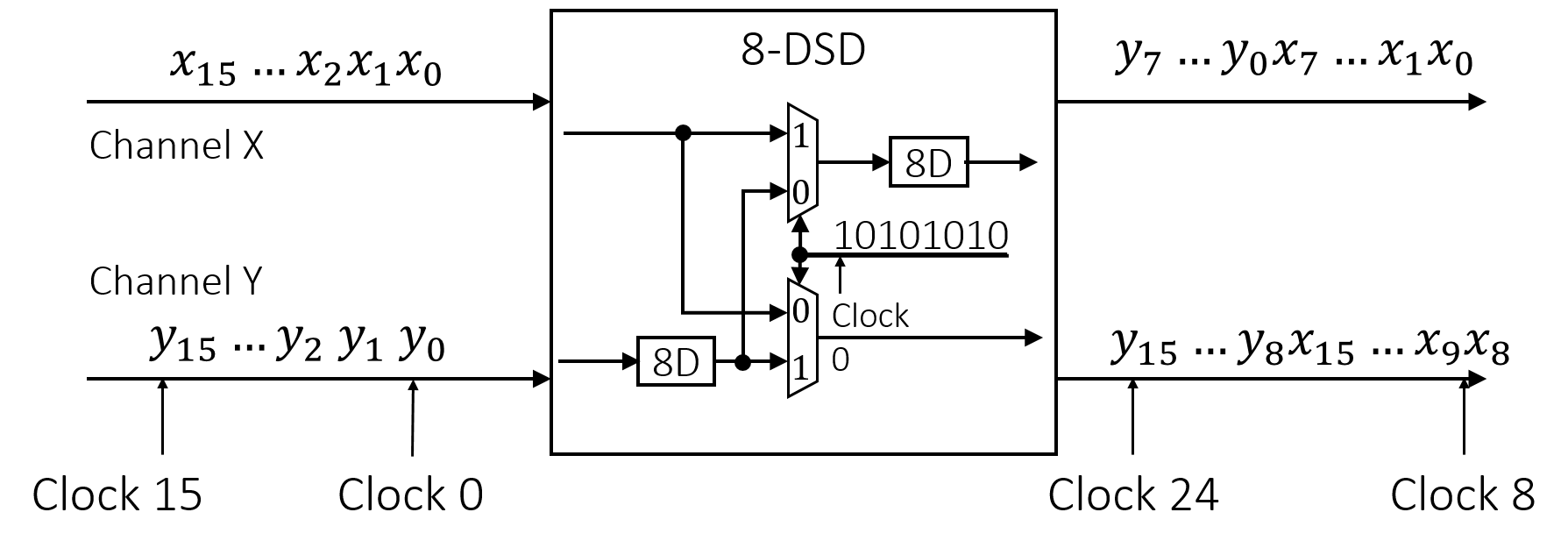}
    \caption{Pre-processing multi-channel input into an interleaved form, where FFT is performed on one channel at a time.}
    \label{fig:8DSD}
\end{figure}

\subsection{Architecture 3: Direct channel interleaving}

Prior architectures perform interleaving on existing architectures with $100\%$ utilization. However, we can derive architecture with full utilization using interleaving even when starting from architectures with half utilization. For example, consider the radix-2 multipath delay commutator (R2MDC) with the folding sets in~\cref{eq:FS_inter_3_null}.

{\scriptsize
\begin{align}
    A = & \left\{A_0, A_1, A_2, A_3, A_4, A_5, A_6, A_7, \emptyset, \emptyset, \emptyset, \emptyset, \emptyset, \emptyset, \emptyset, \emptyset \right\}\nonumber\\
    B = & \left\{\emptyset, \emptyset, \emptyset, \emptyset, B_0, B_1, B_2, B_3, B_4, B_5, B_6, B_7, \emptyset, \emptyset, \emptyset, \emptyset \right\}\nonumber\\
    C = & \left\{\emptyset, \emptyset, \emptyset, \emptyset, \emptyset, \emptyset, C_0, C_1, C_2, C_3, C_4, C_5, C_6, C_7, \emptyset, \emptyset \right\}\nonumber\\
    D = & \left\{\emptyset, \emptyset, \emptyset, \emptyset, \emptyset, \emptyset, \emptyset, D_0, D_1, D_2, D_3, D_4, D_5, D_6, D_7, \emptyset \right\}\label{eq:FS_inter_3_null}
\end{align}
}
There has been extensive research on using different techniques like 2-parallel approaches serial commutators to overcome these inefficiencies. However, we can exploit these null operations in the folding sets to perform interleaving as seen in the folding sets in~\cref{eq:FS_inter_3}.  

{\scriptsize
\begin{align}
    A = & \left\{A_0, A_1, A_2, A_3, A_4, A_5, A_6, A_7, A'_0, A'_1, A'_2, A'_3, A'_4, A'_5, A'_6, A'_7 \right\}\nonumber\\
    B = & \left\{B'_4, B'_5, B'_6, B'_7, B_0, B_1, B_2, B_3, B_4, B_5, B_6, B_7, B'_0, B'_1, B'_2, B'_3 \right\}\nonumber\\
    C = & \left\{ C'_2, C'_3, C'_4, C'_5, C'_6, C'_7, C_0, C_1, C_2, C_3, C_4, C_5, C_6, C_7, C'_0, C'_1 \right\}\nonumber\\
    D = & \left\{D'_1, D'_2, D'_3, D'_4, D'_5, D'_6, D'_7, D_0, D_1, D_2, D_3, D_4, D_5, D_6, D_7, D'_0 \right\}\label{eq:FS_inter_3}
\end{align}
}
The advantage of the proposed folding set is that it dramatically simplifies the pre-processing step. For example, we can use an 8-DSD circuit as shown in \cref{fig:8DSD} to interleave the two channels while also aligning the inputs for the first butterfly stage. Then, after performing the FFT, we can post-process the outputs with another 8-DSD circuit to realign the two channels. The overall architecture showing the pre-processing, post-processing, and FFT modules is shown in~\cref{fig:fft_inter_comm}.
This architecture has further advantages in that it also reduces the complexity to bring the result back into a natural order. As the outputs of the circuit are in a bit-reversed order for only half the FFT size, we use half-size bit reversal circuits to bring them to a natural order. This reversal circuit corresponds to 3 registers per output channel for a 16 point FFT. 

\subsection{Extension to powers of 2 channels}

\begin{figure}
    \centering
    \includegraphics[width=0.8\linewidth]{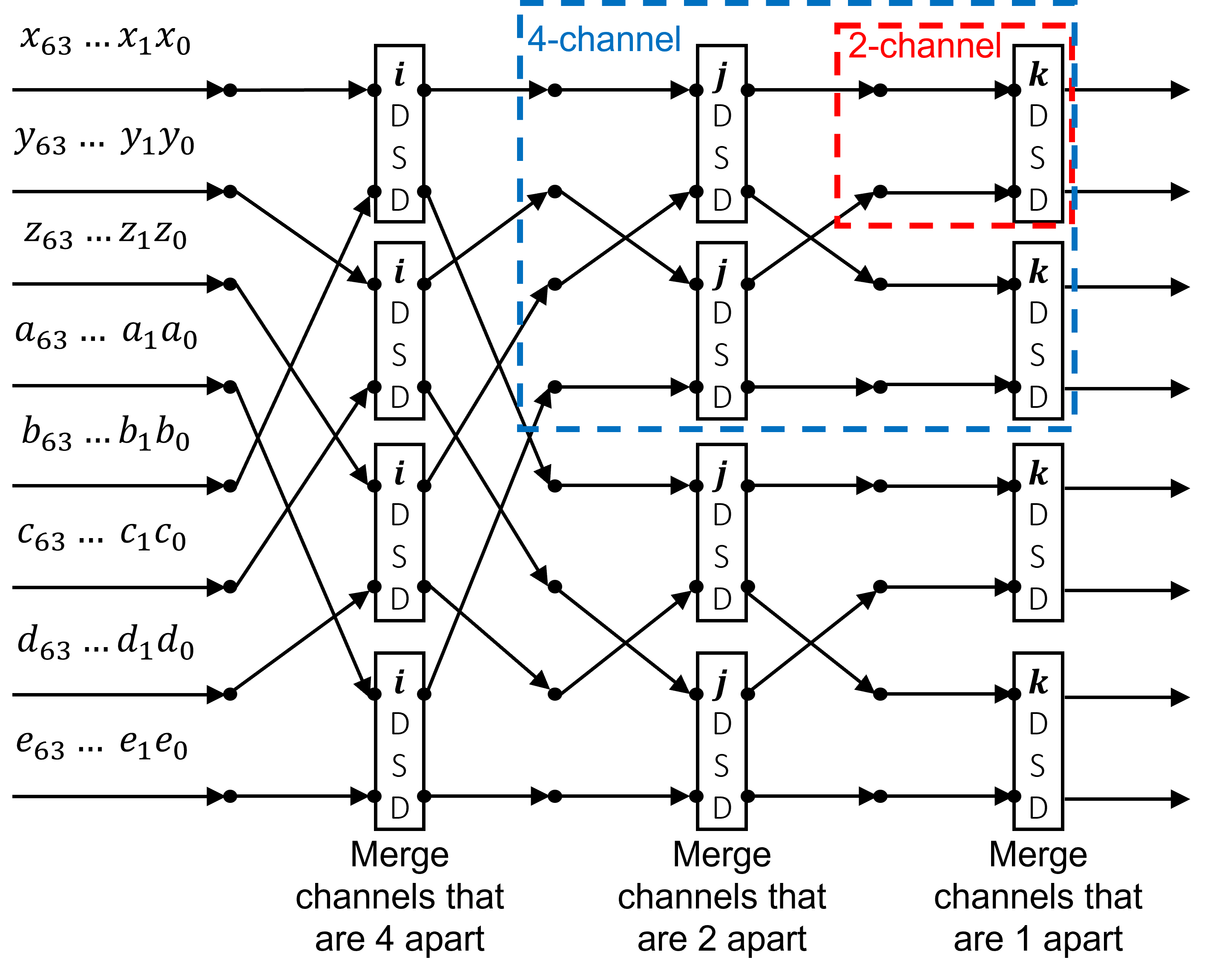}
    \caption{Common circuitry to interleave 2, 4, or 8 channels. The values of $i$, $j$, and $k$ depend on whether we are processing architecture 1 or 3. }
    \label{fig:multi-channel}
\end{figure}

The previous subsections described the process for interleaving two channels. However, this approach can be generalized to any number of channels that is a power of 2. \cref{fig:multi-channel} shows the proposed interleaving pre-processing circuit for Architectures 1 and 3. This is a generalized circuit for 2, 4 and 8 channels. However, adding additional stages can easily extend this approach to more channels. For Architecture 1, the values of the registers $i$, $j$, and $k$ are 4, 2, and 1, respectively. 
In this example, an 8-channel interleaver, the output is directly in the form required for an 8-parallel FFT circuit. For Architecture 3, with a 16-point FFT, the values for i, j, and k are 8, 4, and 2, respectively. For the 64-Point example shown in \cref{fig:multi-channel}, Architecture 3 would require $i$, $j$, and $k$ to be 32, 16, and 8, respectively. 

%% file: sections/S4_comparisons.tex
\section{Comparison of architectures}

All three proposed architectures have identical throughput and processors. Therefore, we compare the memory footprint of these approaches and focus primarily on the pre-processing, post-processing, and reorder steps. These components are the differentiating factors when interleaving multiple channels. 
The pre-processing step interleaves the multiple channels as required by the architecture. The post-processing separates the results into their corresponding channels. The output after post-processing is not in the natural order and requires reordering. The details for each of these components are described in their respective architectures. 
\cref{tab:archs} shows the breakdown of the memory footprint into pre-processing, post-processing, architecture, and reordering registers. 

\begin{table}[tbh]
\caption{Comparison of the memory footprint, in terms of number of registers, of the three proposed architectures on a 16-point DIF FFT}
\label{tab:archs}
\begin{tabular}{|l|c|c|c|c|}
\hline
Architecture & Pre-processing & FFT & Post-processing & Reordering \\
\hline
\hline
Arch 1 & 17 & 28 & 2 & 14\\
\hline
Arch 2 & 18 & 28 & 2 & 18\\
\hline
Arch 3 & 16 & 14 & 16 & 6\\
\hline
\end{tabular}%
\end{table}

\cref{tab:archs} shows that each architecture focuses on different aspects of the design for optimization. We have combined the pre-processing step with the first stage data reordering as these reported together in some architectures for a fair comparison.
Architecture 2 is a direct implementation of interleaving while maintaining the order of computations. However, to bring the inputs to the required order requires complicated pre-processing step that consists of multiple REOCs.
Architecture 1 provides elementary pre-processing and post-processing steps and keeps the base 2-parallel FFT block, that is optimized for performance and output reordering, relatively intact. As a result, Architecture 1 maintains the advantages of the base architecture performing better than Architecture 2. Architecture 3 has the same advantages of Architecture 1 while also simplifying the final reorder circuit. The post-processing step for Architecture 3 un-interleaves the channels and performs half the movement required for the data reordering. Thus we only need an N/2 REOC circuit to perform reordering, where N is the size of the FFT. 

\begin{table}[tbh]
\caption{Comparison of the proposed interleaving pre-processing step versus existing approaches}
\label{tab:all_comparisons}
\begin{tabular}{|p{26mm}|p{12mm}|p{12mm}|p{21mm}|}
\hline
Pre-possessing circuit & Memory elements & Latency & Control complexity\\
\hline
\hline
Memory banks \cite{MIMO_FFT_4, MIMO_FFT_5, MIMO_FFT_6} & $M\times N$ & N & Complex memory reads\\
\hline
Multi-channel commutators \cite{MIMO_FFT_comm_1, MIMO_FFT_comm_2,MIMO_FFT_comm_3} & $(M-1) \times N $ & $(M-1) \times N /M $ & Complex pre-processing control\\
\hline
Proposed architecture 3 & $(M-1) \times N $ &   $(M-1) \times N /M $ & Counter based pre-processing control\\
\hline
\end{tabular}%
\end{table}

\cref{tab:all_comparisons} shows a detailed comparison of the proposed pre-processing method with existing literature. Here, $M$ is the number of interleaved channels, and $N$ is the size of the FFT. The prior approaches are based on using a large memory bank \cite{MIMO_FFT_4, MIMO_FFT_5, MIMO_FFT_6} to store the information from all channels and retrieving the information as required. However, this method often employs complicated memory access patterns to retrieve the data in the required form. Additionally, they have a larger memory footprint as they do not optimize register utilization by performing a lifetime analysis. 

Multi-channel commutators are increasingly common in MIMO-OFDM applications \cite{MIMO_FFT_comm_1, MIMO_FFT_comm_2,MIMO_FFT_comm_3}, and they overcome some of the shortcomings of memory bank approaches. This method minimizes the number of registers required to pre-process the inputs and have optimal latency. However, existing approaches are ad-hoc and do not design generalized structures for various channels and points. Additionally, the control circuit complexity of the pre-processing step increases with the number of channels, $M$, and FFT size, $N$.


%% file: sections/S5_conclusion.tex
\section{Conclusion}

This paper presented a novel approach to design multi-channel interleaved architectures from existing FFT architectures systematically. 
The paper proposed three separate approaches to deriving pipeline models based on folding and interleaving. In addition, the paper presents a framework for efficiently interleaving any number of channels with elementary control logic. Thus the paper can be used as a baseline to deriving any $C$ channel FFT module from a $C$ parallel baseline FFT architecture.